\input harvmac
\input epsf

\newcount\figno
\figno=0
\def\fig#1#2#3{
\par\begingroup\parindent=0pt\leftskip=1cm\rightskip=1cm\parindent=0pt
\baselineskip=11pt
\global\advance\figno by 1
\midinsert
\epsfxsize=#3
\centerline{\epsfbox{#2}}
\vskip 12pt
{\bf Fig.\ \the\figno: } #1\par
\endinsert\endgroup\par
}
\def\figlabel#1{\xdef#1{\the\figno}}

\def\sh{\mathop{\rm sh}\nolimits}
\def\ch{\mathop{\rm ch}\nolimits}

\lref\ZakharovPP{
  V.~E.~Zakharov and A.~V.~Mikhailov,
  ``Relativistically Invariant Two-Dimensional Models In Field Theory
  Integrable By The Inverse Problem Technique. (In Russian),''
  Sov.\ Phys.\ JETP {\bf 47}, 1017 (1978)
  [Zh.\ Eksp.\ Teor.\ Fiz.\  {\bf 74}, 1953 (1978)].
}

\lref\DeVegaXC{
  H.~J.~De Vega and N.~G.~Sanchez,
  ``Exact Integrability Of Strings In D-Dimensional De Sitter Space-Time,''
  Phys.\ Rev.\  D {\bf 47}, 3394 (1993).
}

\lref\HarnadWE{
  J.~P.~Harnad, Y.~Saint Aubin and S.~Shnider,
  ``B\"acklund Transformations For Nonlinear Sigma Models With Values In
  Riemannian Symmetric Spaces,''
  Commun.\ Math.\ Phys.\  {\bf 92}, 329 (1984).
}

\lref\CombesRW{
  F.~Combes, H.~J.~de Vega, A.~V.~Mikhailov and N.~G.~Sanchez,
  ``Multistring solutions by soliton methods in de Sitter space-time,''
  Phys.\ Rev.\  D {\bf 50}, 2754 (1994)
  [arXiv:hep-th/9310073].
}

\lref\ReyIK{
  S.~J.~Rey and J.~T.~Yee,
  ``Macroscopic strings as heavy quarks in large $N$ gauge theory and anti-de
  Sitter supergravity,''
  Eur.\ Phys.\ J.\  C {\bf 22}, 379 (2001)
  [arXiv:hep-th/9803001].
}

\lref\MaldacenaIM{
  J.~M.~Maldacena,
  ``Wilson loops in large $N$ field theories,''
  Phys.\ Rev.\ Lett.\  {\bf 80}, 4859 (1998)
  [arXiv:hep-th/9803002].
}

\lref\GubserTV{
  S.~S.~Gubser, I.~R.~Klebanov and A.~M.~Polyakov,
  ``A semi-classical limit of the gauge/string correspondence,''
  Nucl.\ Phys.\  B {\bf 636}, 99 (2002)
  [arXiv:hep-th/0204051].
}

\lref\KruczenskiFB{
  M.~Kruczenski,
  ``A note on twist two operators in ${\cal N} = 4$
  SYM and Wilson loops in Minkowski signature,''
  JHEP {\bf 0212}, 024 (2002)
  [arXiv:hep-th/0210115].
}

\lref\AnastasiouKJ{
  C.~Anastasiou, Z.~Bern, L.~J.~Dixon and D.~A.~Kosower,
  ``Planar amplitudes in maximally supersymmetric Yang-Mills theory,''
  Phys.\ Rev.\ Lett.\  {\bf 91}, 251602 (2003)
  [arXiv:hep-th/0309040].
}

\lref\TseytlinCJ{
  A.~A.~Tseytlin,
  ``Semiclassical strings in $AdS_5 \times S^5$ and scalar operators in
  ${\cal N} = 4$
  SYM theory,''
  Comptes Rendus Physique {\bf 5}, 1049 (2004)
  [arXiv:hep-th/0407218].
}

\lref\TseytlinXA{
  A.~A.~Tseytlin,
  ``Semiclassical strings and AdS/CFT,''
  arXiv:hep-th/0409296.
}

\lref\ArutyunovYX{
  G.~Arutyunov and S.~Frolov,
  ``Integrable Hamiltonian for classical strings on $AdS_5 \times S^5$,''
  JHEP {\bf 0502}, 059 (2005)
  [arXiv:hep-th/0411089].
}

\lref\BernIZ{
  Z.~Bern, L.~J.~Dixon and V.~A.~Smirnov,
  ``Iteration of planar amplitudes in maximally supersymmetric Yang-Mills
  theory at three loops and beyond,''
  Phys.\ Rev.\  D {\bf 72}, 085001 (2005)
  [arXiv:hep-th/0505205].
}

\lref\PlefkaBK{
  J.~Plefka,
  ``Spinning strings and integrable spin chains in the AdS/CFT
  correspondence,''
  arXiv:hep-th/0507136.
}

\lref\HofmanXT{
  D.~M.~Hofman and J.~M.~Maldacena,
  ``Giant magnons,''
  J.\ Phys.\ A  {\bf 39}, 13095 (2006)
  [arXiv:hep-th/0604135].
}

\lref\DoreyDQ{
  N.~Dorey,
  ``Magnon bound states and the AdS/CFT correspondence,''
  J.\ Phys.\ A  {\bf 39}, 13119 (2006)
  [arXiv:hep-th/0604175].
}

\lref\ChenGE{
  H.~Y.~Chen, N.~Dorey and K.~Okamura,
  ``Dyonic giant magnons,''
  JHEP {\bf 0609}, 024 (2006)
  [arXiv:hep-th/0605155].
}

\lref\SpradlinWK{
  M.~Spradlin and A.~Volovich,
  ``Dressing the giant magnon,''
  JHEP {\bf 0610}, 012 (2006)
  [arXiv:hep-th/0607009].
}

\lref\KalousiosXY{
  C.~Kalousios, M.~Spradlin and A.~Volovich,
  ``Dressing the giant magnon. II,''
  JHEP {\bf 0703}, 020 (2007)
  [arXiv:hep-th/0611033].
}

\lref\BernEW{
  Z.~Bern, M.~Czakon, L.~J.~Dixon, D.~A.~Kosower and V.~A.~Smirnov,
  ``The Four-Loop Planar Amplitude and Cusp Anomalous Dimension in Maximally
  Supersymmetric Yang-Mills Theory,''
  Phys.\ Rev.\  D {\bf 75}, 085010 (2007)
  [arXiv:hep-th/0610248].
}

\lref\CachazoAZ{
  F.~Cachazo, M.~Spradlin and A.~Volovich,
  ``Four-Loop Cusp Anomalous Dimension From Obstructions,''
  Phys.\ Rev.\  D {\bf 75}, 105011 (2007)
  [arXiv:hep-th/0612309].
}

\lref\AldayHR{
  L.~F.~Alday and J.~Maldacena,
  ``Gluon scattering amplitudes at strong coupling,''
  JHEP {\bf 0706}, 064 (2007)
  [arXiv:0705.0303 [hep-th]].
}

\lref\CachazoAD{
  F.~Cachazo, M.~Spradlin and A.~Volovich,
  ``Four-Loop Collinear Anomalous Dimension in ${\cal N} = 4$
  Yang-Mills Theory,''
  arXiv:0707.1903 [hep-th].
}

\lref\KruczenskiCY{
  M.~Kruczenski, R.~Roiban, A.~Tirziu and A.~A.~Tseytlin,
  ``Strong-coupling expansion of cusp anomaly and gluon amplitudes from quantum
  open strings in $AdS_5 \times S^5$,''
  arXiv:0707.4254 [hep-th].
}

\Title
{\vbox{
\baselineskip12pt
\hbox{Brown-HET-1486}
}}
{\vbox{
\centerline{Dressing the Giant Gluon}
}}

\centerline{
Antal Jevicki, Chrysostomos Kalousios, Marcus Spradlin and Anastasia Volovich
}

\vskip .5in
\centerline{Brown University}
\centerline{Providence, Rhode Island 02912 USA}

\vskip .5in
\centerline{\bf Abstract}

We demonstrate the applicability of the dressing method to the
problem of constructing new
classical solutions for Euclidean worldsheets
in anti-de Sitter space.
The motivation stems from recent work of
Alday and Maldacena, who
studied gluon scattering amplitudes at strong coupling 
using a generalization of a particular worldsheet found by
Kruczenski
whose edge
traces a path composed of light-light segments on the boundary of $AdS$.
We dress this `giant gluon'
to find new solutions in $AdS_3$ and $AdS_5$ whose
edges trace out more complicated, timelike curves on the boundary.
These solutions may be used to calculate
certain Wilson loops via AdS/CFT.

\Date{August 2007}

\listtoc
\writetoc

\newsec{Introduction}

Classical string solutions play an important role in
exploring the AdS/CFT
correspondence
(see~\GubserTV\ and~\refs{\TseytlinCJ,\TseytlinXA,\PlefkaBK} for reviews).
Generally speaking such
solutions fall into two categories.  On the one hand there are
closed string energy eigenstates in $AdS$, which are in correspondence
with gauge invariant operators of definite scaling dimension
in the dual gauge theory.
On the other hand we can also consider open strings
which end along some curve on
the boundary of $AdS$, corresponding to Wilson
loops~\refs{\ReyIK,\MaldacenaIM}.

An important example of the former is the so-called `giant
magnon' of Hofman and Maldacena~\HofmanXT, which is dual
to a single elementary excitation in the gauge theory picture.
More general states containing arbitrary numbers of bound or scattering
states of magnons correspond to more general classical string
solutions~\refs{\DoreyDQ,\ChenGE,\SpradlinWK}.
These solutions can be constructed algebraically using the
dressing method~\refs{\SpradlinWK,\KalousiosXY}, a well-known
technique~\refs{\ZakharovPP,\HarnadWE} for
generating solutions of classically integrable equations.

In this paper we turn our attention to the latter, demonstrating
the applicability of the dressing method to the problem of
constructing certain new Euclidean minimal area
surfaces in anti-de Sitter space\foot{The dressing method has also
been used to construct Minkowskian worldsheets in de Sitter
space~\refs{\DeVegaXC,\CombesRW}.}.
To apply the dressing method it is necessary to choose
some solution of the classical equations of motion to use as the
`vacuum', which is then `dressed' to build more general solutions.
For the giant magnon system considered
in~\refs{\SpradlinWK,\KalousiosXY} it was natural to choose as
vacuum the solution describing a pointlike string moving at the
speed of light around the equator of the $S^5$, since this state
corresponds to the natural vacuum in the spin chain picture.

For the present problem we choose as vacuum a particular solution,
shown in Figure~1,
originally used by Kruczenski~\KruczenskiFB\ to study the cusp
anomalous dimension via AdS/CFT.
It is the minimal area surface which meets the boundary of global $AdS_3$
along four intersecting light-like lines.
This solution was recently
generalized, and given a new interpretation,
by Alday and Maldacena~\AldayHR, who gave a prescription
for computing planar gluon scattering amplitudes
in ${\cal N} = 4$ Yang-Mills
at strong coupling
using the AdS/CFT correspondence and found perfect agreement with
the structure predicted on the basis of previously conjectured
iteration relations for perturbative
multiloop gluon
amplitudes~\refs{\AnastasiouKJ,\BernIZ,\BernEW,\CachazoAZ,\CachazoAD}.

The Alday-Maldacena 
prescription is (classically)
computationally equivalent to the
problem of evaluating a Wilson loop 
composed of light-like segments.  According to the AdS/CFT
dictionary, such a Wilson loop is computed by evaluating the area
of the surface in Figure~1.
The interpretation of this surface in terms of a gluon scattering
process suggests calling
this kind of solution a `giant gluon.'

\fig{The `giant gluon' solution (3.1) in $AdS_3$ global coordinates.
The gluons
follow the four light-like segments on the boundary of $AdS_3$ where
the worldsheet ends.
}{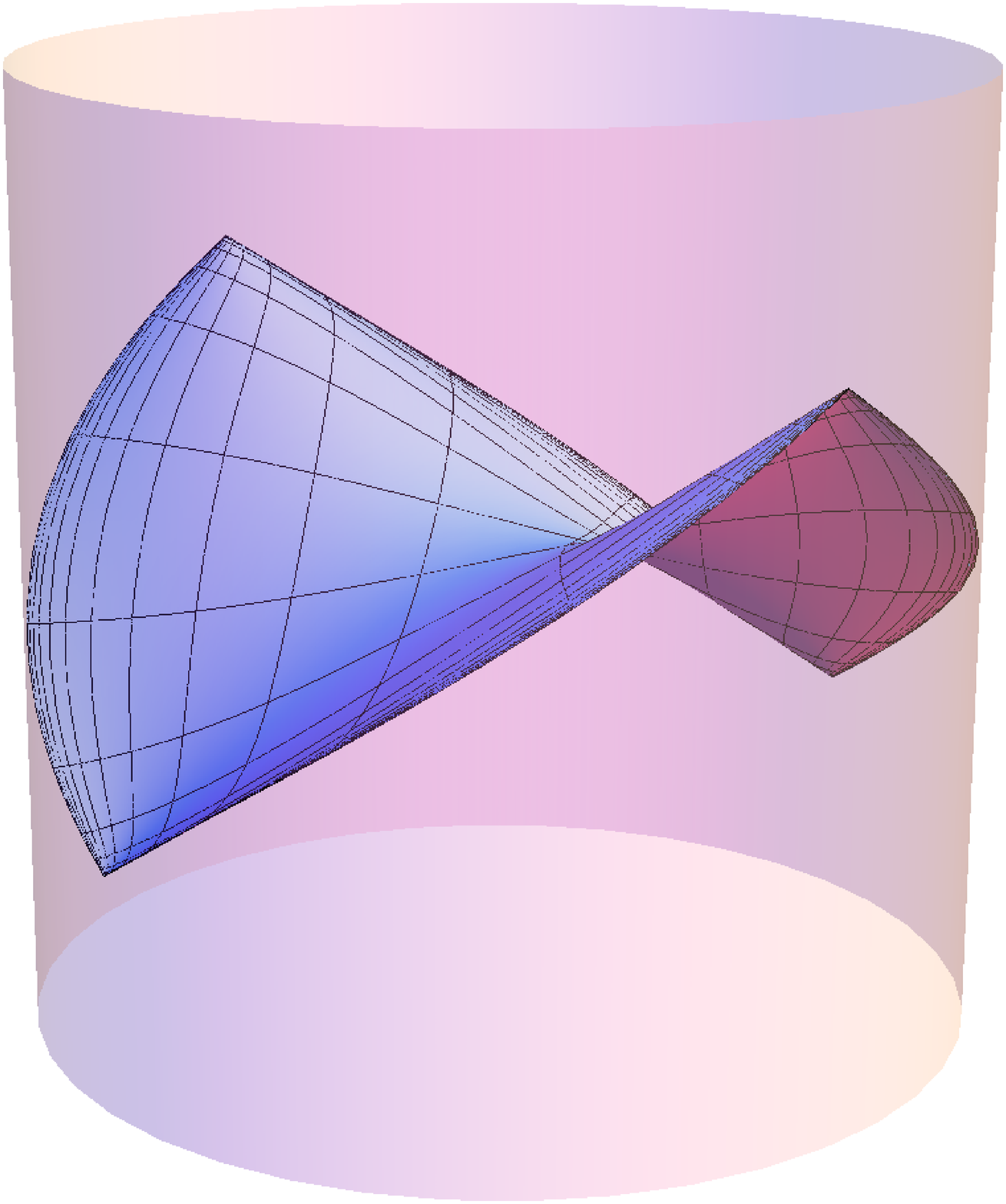}{59.972mm}

We dress the giant gluon to find new minimal area
surfaces in $AdS_3$ and $AdS_5$ whose
edges trace out more complicated, timelike curves on the boundary of AdS.
It is not clear whether these new solutions have any interpretation
as a scattering process of the type studied in~\AldayHR, although they
do have straightforward interpretations in terms of Wilson loops.
However, when calculating a Wilson loop one usually first
specifies a curve on the boundary of AdS and then finds the minimal
area surface bounding that curve.
In contrast, the dressing method provides the minimal area
surface without telling us the curve that it spans, i.e. without
telling us which Wilson loop it is calculating.  That information
must be read off directly by analyzing the solution to see
where it reaches the boundary of $AdS$, a procedure that we will
see is rather nontrivial.

The outline of this paper is as follows.
In section 2 we demonstrate the applicability of the dressing
method, focusing on the $AdS_3$ case which is simpler because there the
problem can be mapped into the SU(1,1) principal chiral model.
In section 3 we discuss the dressed giant gluon in $AdS_3$, display explicit
formulas for a special case of the solution, and analyze in detail the
edge of the worldsheet on the boundary of $AdS_3$.  In section 4 we turn
to the more complicated construction for $AdS_5$ solutions using
the SU(2,2)/SO(4,1) coset model, and present some examples.

The main goal of this paper is to demonstrate the applicability of the
dressing method.  Although we consider a few examples, they amount
to only a small subset of the simplest possible solutions.  It would
be very interesting to more fully explore the parameter space of
solutions that can be obtained.
It would also be interesting to evaluate the (regulated) areas of
these solutions, thereby calculating the corresponding Wilson loops in
gauge theory.
The giant gluon shown in Figure~1 can actually be
related~\KruczenskiCY,
by analytic continuation and a conformal transformation, to a
closed string energy eigenstate (a limit of the GKP spinning string~\GubserTV).
It would be interesting to see whether it is possible to relate
more general Euclidean worldsheets of the type we consider to
various closed string states.

\newsec{$AdS$ Dressing Method}

The dressing method~\ZakharovPP\ is a general technique for
constructing solutions of classically integrable equations.
As we review shortly,
at the heart of the method lies the ability to transform
nonlinear equations of motion into a linear system for
an auxiliary field.
Here we apply this very general method to the specific problem of
constructing minimal area Euclidean
worldsheets in anti de-Sitter space.
Initially we restrict our attention to $AdS_3$, where the problem
relates to the SU(1,1) principal chiral model, deferring the slightly
more complicated $AdS_5$ case to section 4.
Many of the equations in this section are similar to those
appearing in~\refs{\SpradlinWK,\KalousiosXY}, which the reader may consult for
further details.
The two most significant differences
compared to the SU(2) principal chiral model considered
in~\SpradlinWK\ are that we
use complex coordinates $z, \bar{z}$ on the worldsheet,
which is now Euclidean, and
that the indefinite SU(1,1) metric
significantly changes the behavior of the solutions compared to SU(2).

We parameterize
AdS${}_d$ with $d+1$ embedding coordinates $\vec{Y}$
subject to the constraint
\eqn\aaa{
\vec{Y} \cdot \vec{Y} \equiv - Y_{-1}^2 - Y_{0}^2
+ Y_{1}^2 + Y_{2}^2 + \cdots + Y_{d-1}^2 = - 1.
}
Minimal area worldsheets are given by solutions to the conformal
gauge equations of motion
\eqn\eom{
\partial \bar{\partial} \vec{Y} - \vec{Y} \left( \partial \vec{Y}
\cdot \bar{\partial} \vec{Y} \right) = 0
}
subject to the Virasoro constraints
\eqn\virasoro{
\partial \vec{Y} \cdot \partial \vec{Y} =
\bar{\partial} \vec{Y} \cdot \bar{\partial} \vec{Y} = 0.
}
Here and throughout the paper we use complex coordinates
\eqn\aaa{
z = {1 \over 2} (u_1 + i u_2), \qquad
\bar{z} = {1 \over 2} (u_1 - i u_2),
}
with
\eqn\aaa{
\partial = \partial_1 - i \partial_2, \qquad \bar{\partial} = \partial_1
+ i \partial_2.
}

Our first step is to recast the system \eom, \virasoro\ 
into the form of a principal chiral model for a matrix-valued field
$g$ satisfying the equation of motion
\eqn\pcmeom{
\bar{\partial} A + \partial \bar{A} = 0
}
in terms of the currents
\eqn\Adef{
A = i \partial g\, g^{-1}, \qquad
\bar{A} = i \bar{\partial} g\, g^{-1}.
}
Note that the relation
\eqn\automatic{
\bar{\partial} A - \partial \bar{A} - i [A, \bar{A}] = 0
}
follows automatically from \Adef.

To see how this is done let us consider
for simplicity first
the $AdS_3$ case.
Here we use the coordinates $\vec{Y}$ to parameterize
an element $g$ of SU(1,1) according to
\eqn\parametrization{
g = \pmatrix{ Z_1 & Z_2 \cr
\bar{Z}_2 & \bar{Z}_1 }, \qquad
Z_1 = Y_{-1} + i Y_0, \qquad Z_2 = Y_1 + i Y_2,
}
which satisfies
\eqn\aaa{
g^\dagger M g = M, \qquad M = \pmatrix{+1 & 0 \cr 0 & -1}
}
and
\eqn\aaa{
\det g = - \vec{Y} \cdot \vec{Y} = +1.
}
It is easy to check that the systems \eom, \virasoro\ and
\pcmeom, \automatic\ are equivalent to each other under this
change of variables.

Next we transform the nonlinear second-order
system \pcmeom, \Adef\ for $g(z,\bar{z})$
into a linear, first-order system for
an auxiliary field $\Psi(z,\bar{z},\lambda)$
at the expense of introducing a new
complex parameter $\lambda$ called the spectral parameter.
Specifically, the two equations \pcmeom, \Adef\ are equivalent to
\eqn\linear{
i \partial \Psi = {A \Psi \over 1 + i \lambda}, \qquad
i \bar{\partial} \Psi = {\bar{A} \Psi \over 1 - i \lambda}.
}
For later convenience we have
rescaled our definition of $\lambda$ in this equation
by a factor of $i$ compared
to the conventions of~\refs{\SpradlinWK,\KalousiosXY}.

To apply the dressing method we begin with any known solution $g$
(which we refer to as the `vacuum' for the dressing method, though
we emphasize that any 
solution may be chosen as the vacuum) and then solve the linear system
\linear\ to find $\Psi(\lambda)$ subject to the initial condition
\eqn\fff{
\Psi(\lambda = 0) = g.
}
In addition we impose on $\Psi(\lambda)$ the SU(1, 1) conditions
\eqn\unitarity{
\Psi^\dagger(\bar{\lambda}) M \Psi(\lambda) = M, \qquad
\det \Psi(\lambda) = 1.
}
The purpose of the factor of $i$ mentioned below \linear\ is to
avoid the need to take $- \bar{\lambda}$ instead of $\bar{\lambda}$ in the
first relation here.

Then we make a `gauge transformation' of the form
\eqn\psiprime{
\Psi'(\lambda) = \chi(\lambda) \Psi(\lambda).
}
If $\chi(\lambda)$ were independent of $z$ and $\bar{z}$ this would
be an uninteresting SU(1,1)
gauge transformation.  Instead we want $\chi(\lambda)$
to depend on $z$ and $\bar{z}$ but in such a way that $\Psi'(\lambda)$
continues to satisfy \linear\ and hence $\Psi'(0)$ provides
a new solution to \pcmeom, \automatic.
For $AdS_3$ it is not
hard to show that
this is accomplished by taking $\chi(\lambda)$ to have the
form
\eqn\chiadsthree{
\chi(\lambda) = 1 + {\lambda_1 - \bar{\lambda}_1 \over
\lambda - \lambda_1} P
}
where $\lambda_1$ is an arbitrary complex parameter
and $P$ is a projection operator onto any vector of the form
$v_1 \equiv \Psi(\bar{\lambda}_1) v$ for any constant vector $v$.
Concretely, $P$ is therefore given by
\eqn\projone{
P = { v_1 v_1^\dagger M \over v_1^\dagger M v_1}.
}
As in~\SpradlinWK\ there is a minor remaining detail that~\chiadsthree\ has
\eqn\aaa{
\det \chi(\lambda) = \bar{\lambda}_1/\lambda_1
}
so in order for $g'$ to lie in SU(1,1) rather than U(1,1)
we should rescale $g'$ by the constant
phase factor $\sqrt{\lambda_1/\bar{\lambda}_1}$ to ensure that
it has unit determinant.
To summarize, the desired dressed solution is given by
\eqn\gprime{
g' = \sqrt{\lambda_1 \over \bar{\lambda}_1} \left[
1 + { \lambda_1 - \bar{\lambda}_1 \over - \lambda_1} P \right] \Psi(0).
}
The real embedding coordinates $\vec{Y}'$ of the dressed solution
may then
be read off from $g'$ using the parameterization~\parametrization.
The resulting solution is characterized by the complex
parameter $\lambda_1$ and the choice of the constant vector $v$.

\newsec{$AdS_3$ Solutions}

In this section we obtain new solutions for worldsheets in $AdS_3$
via the dressing method, taking as `vacuum' the giant gluon
solution~\refs{\KruczenskiFB,\AldayHR}
\eqn\solone{
\vec{Y} = \pmatrix{
Y_{-1} \cr
Y_0 \cr
Y_1 \cr
Y_2 }
=
\pmatrix{
\cosh u_1 \cosh u_2 \cr
\sinh u_1 \sinh u_2 \cr
\sinh u_1 \cosh u_2 \cr
\cosh u_1 \sinh u_2}.
}
Using the $AdS_3$ parameterization~\parametrization\ we find
from \Adef\ that
\eqn\aaa{
\eqalign{
A &= 2 \pmatrix{
-\cosh u_2 \sinh u_2 &  i \cosh^2 u_2 \cr
 i \sinh^2 u_2 & +\cosh u_2 \sinh u_2},\cr
\bar{A} &= 2 \pmatrix{
- \cosh u_2 \sinh u_2 &  i \sinh^2 u_2 \cr
 i \cosh^2 u_2 & +\cosh u_2 \sinh u_2}.
}}
Then
a solution to the linear system \linear\ for $\Psi(\lambda)$ is\foot{We
will occasionally use $\sh,\ch$ instead of $\sinh,\cosh$ to compactify
otherwise lengthy formulas.}
\eqn\eee{
\Psi(\lambda)=
\pmatrix{
m_- \ch Z \ch u_2 + i m_+ \sh Z \sh u_2 &
m_- \sh Z \ch u_2 + i m_+ \ch Z \sh u_2 \cr
m_+ \sh Z \ch u_2 - i m_- \ch Z \sh u_2 &
m_+ \ch Z \ch u_2 - i m_- \sh Z \sh u_2}
}
where
\eqn\aaa{
m_+ = 1/m_- = \left({1 + i \lambda \over 1 - i \lambda}\right)^{1/4}, \qquad
Z = m_-^2 z + m_+^2 \bar{z}.
}
The solution \eee\ has been chosen to satisfy
the desired constraints
\unitarity\ as well as the initial condition
\eqn\aaa{
\Psi(0) = \pmatrix{
\cosh u_1 \cosh u_2 + i \sinh u_1 \sinh u_2 &
\sinh u_1 \cosh u_2 + i \cosh u_1 \sinh u_2 \cr
\sinh u_1 \cosh u_2 - i \cosh u_1 \sinh u_2 &
\cosh u_1 \cosh u_2 - i \sinh u_1 \sinh u_2 
},
}
correctly reproducing the giant gluon solution~\solone\ embedded
into SU(1,1) according to~\parametrization.
The dressed solution $g'$ is then given by~\gprime.

\subsec{A special case}

Since the general solution is rather complicated, we present here
an explicit formula for the dressed solution for the
particular choice of initial vector $v = \pmatrix{1 & i}$, with
$\lambda_1$ arbitrary.
We find that the dressed SU(1,1) principal chiral field takes
the form
\eqn\guuu{
g' = \pmatrix{Z'_1 & Z'_2 \cr
\bar{Z}'_2 & \bar{Z}'_1}
}
where
\eqn\zs{
Z'_1 = {1 \over |\lambda_1|} {\vec{Y} \cdot \vec{N}_1 \over D}, \qquad Z'_2 =
{1 \over |\lambda_1|} {\vec{Y} \cdot \vec{N}_2 \over D}
}
in terms of the numerator factors
\eqn\dressone{\eqalign{
\vec{N}_1 &=
 \pmatrix{- (\bar{\lambda}_1 |m|^2 - \lambda_1) \cosh(Z + \bar{Z})
+ i (\bar{\lambda}_1 |m|^2 + \lambda_1) \sinh(Z - \bar{Z}) \cr
- (\lambda_1 |m|^2 + \bar{\lambda}_1) \sinh(Z - \bar{Z}) - i (\lambda_1 |m|^2
- \bar{\lambda}_1) \cosh(Z + \bar{Z}) \cr
(\lambda_1 - \bar{\lambda}_1) \bar{m} (\sinh(Z + \bar{Z}) - i
\cosh(Z - \bar{Z})) \cr
(\lambda_1 - \bar{\lambda}_1) m ( \cosh(Z - \bar{Z}) - i
\sinh(Z + \bar{Z}))},\cr
\vec{N}_2 &=  \pmatrix{
- (\lambda_1 - \bar{\lambda}_1) \bar{m} ( \sinh(Z + \bar{Z}) - i
\cosh(Z - \bar{Z}) ) \cr
- (\lambda_1 - \bar{\lambda}_1) m ( \cosh(Z - \bar{Z}) - i \sinh(Z + \bar{Z}))\cr
+ (\bar{\lambda}_1 |m|^2 - \lambda_1) \cosh(Z + \bar{Z}) - i (\bar{\lambda}_1
|m|^2 + \lambda_1) \sinh(Z - \bar{Z})\cr
+ (\lambda_1 |m|^2 + \bar{\lambda}_1) \sinh(Z - \bar{Z}) +
i (\lambda_1 |m|^2 - \bar{\lambda}_1) \cosh(Z + \bar{Z})},
}}
$\vec{Y}$ given in \solone,
and the denominator
\eqn\aaa{
D = ( |m|^2 - 1) \cosh(Z + \bar{Z}) - i (|m|^2 + 1)
\sinh(Z - \bar{Z}).
}
In these expressions
\eqn\mmbardefs{
m = \left({1 + i \lambda_1 \over 1 - i \lambda_1}\right)^{1/2}, \qquad
\bar{m} = \left({1 - i \bar{\lambda}_1 \over 1 + i \bar{\lambda}_1}\right)^{1/2},
}
and
\eqn\yyy{
Z = z/m + m \bar{z}, \qquad
\bar{Z} = \bar{z}/\bar{m} + \bar{m} z.
}

The real embedding coordinates  $\vec{Y}'$ of the dressed solution
are easily read off from~\guuu\ using~\parametrization.
In Figure~2 we plot a representative example of the solution~\zs.
However before one can make sense of the plot we must understand
the behavior of~\zs\ at the boundary of AdS, which we address in
the next subsection.

\fig{An example of a surface described by the
solution~\zs\ for the particular choice
$\lambda_1 = 1/2 + i/3$.
}{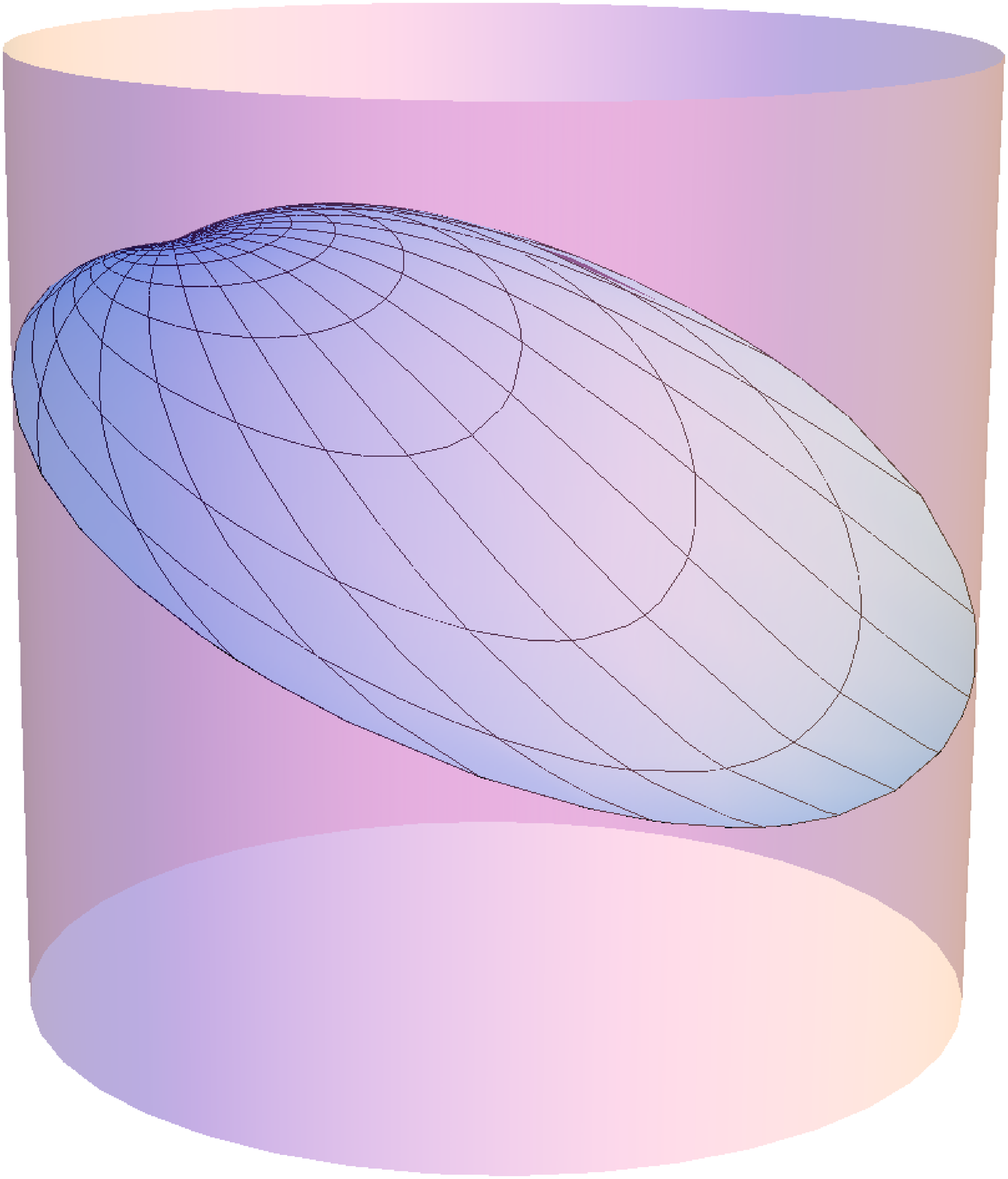}{59.972mm}

\subsec{In search of the Wilson loop}

Minimal area worldsheets in $AdS_5$
are related to Wilson loops in the dual gauge
theory~\refs{\ReyIK,\MaldacenaIM}.  According to the
AdS/CFT dictionary, in order to calculate the expectation
value of the Wilson loop for some
closed path ${\cal C}$ on the boundary of $AdS$ we should first find
the minimal area surface (or surfaces) in $AdS$ which spans that
curve and then calculate $e^{-A}$ where $A$ is the (regulated) area
of the minimal surface.

The solutions we have obtained by the dressing method turn this procedure
on its head.
In the previous subsection we displayed an explicit example of such 
a solution,
which indeed describes a minimal area Euclidean worldsheet in
$AdS_3$,
but it is not immediately clear what the corresponding curve ${\cal C}$
is whose Wilson loop the solution computes.
In order to answer this question we must look at~\zs\ and find
the locus ${\cal C}$ where the worldsheet reaches the boundary
of $AdS_3$---this
will tell us which Wilson loop we are computing.

In global $AdS$ coordinates, the familiar radial coordinate $\rho$ is
related to the coordinates appearing in~\parametrization\ according
to
\eqn\aaa{
\cosh^2 \rho = |Z_1|^2, \qquad \sinh^2 \rho = |Z_2|^2.
}
Hence the boundary of $AdS_3$ lies at $Z_i = \infty$.
Before proceeding with our complicated dressed solution let us pause
to note that the giant gluon solution~\solone\ reaches the boundary
of $AdS_3$ precisely when $|u_1| \to \infty$ or $|u_2| \to \infty$.
Moreover the four `edges' of the worldsheet, at $u_1 \to +\infty$,
$u_1 \to -\infty$, $u_2 \to + \infty$ and $u_2 \to -\infty$,
sit on four separate null lines on the boundary of $AdS_3$ which intersect
each other at four cusps~\refs{\KruczenskiFB,\AldayHR} to form the closed
curve ${\cal C}$.

Looking at the dressed solution
\zs\ we see a feature which makes it significantly more complicated
to understand than the giant gluon.
The presence of the nontrivial denominator factor
\eqn\aaa{
D = (|m|^2 - 1) \cosh(Z+\bar{Z}) - i (|m|^2 + 1) \sinh(Z - \bar{Z})
}
in~\zs\ means
that the solution reaches the boundary of $AdS_3$ any time $D = 0$,
which occurs at finite (rather than infinite) values of
the worldsheet coordinates $z, \bar{z}$.
In fact since $D$ is periodic in $Z$ (with period $\pi i$),
the solution reaches the boundary of $AdS_3$ infinitely many times
as we allow $z$ (and hence $Z$) to vary across the complex plane.
It is important to note that while $D$ is periodic, the full solution
is not.

If we define real variables $U_i$ according to
\eqn\zzz{
Z = (U_1 + i U_2)/2, \qquad \bar{Z} = (U_1 - i U_2)/2
}
then the locus $\tilde{C}$
of points on the worldsheet where the solution reaches
the boundary of $AdS_3$ is
\eqn\aaa{
D = (|m|^2 - 1) \cosh U_1 + (|m|^2 + 1) \sin U_2 = 0.
}
This equation describes an infinite array of oval-shaped curves
$\tilde{\cal C}_j$
periodically distributed along the $U_2$ axis and centered
at $(U_1,U_2) = (0, 2 \pi j + \pi/2)$.
Note that the curves $\tilde{\cal C}_j$ in the worldsheet coordinates
are not to be confused
with their images ${\cal C}_j$ on the boundary of $AdS_3$
under the map~\zs.
In particular the $\tilde{\cal C}_j$ are unphysical artifacts of the
particular coordinate system we happen to be using on the worldsheet---only
the curves ${\cal C}_j$ on the boundary are physically meaningful.

To summarize,
we find that the solution~\zs\ actually describes not one but
infinitely many different minimal area surfaces in $AdS_3$, each spanning
a different curve ${\cal C}_j$ on the boundary.  In order to isolate
any given worldsheet $j$ we restrict the worldsheet coordinates $U_1,U_2$
to range over the interior of the curve $\tilde{\cal C}_j$.
In particular, 
in order to find the area of the $j$-th worldsheet, and hence
calculate the expectation value of the Wilson loop corresponding
to the curve ${\cal C}_j$,
one should integrate the induced volume
element on the worldsheet only over the region $\tilde{\cal C}_j$.
It would be interesting to pursue this calculation further, although
we will not do so here.

\subsec{A very special case}

In the previous subsection we explained that the minimal area
surfaces generated by the dressing method actually calculate
infinitely many different Wilson loops.  In general the solutions
are sufficiently complicated that we find it necessary to analyze them
numerically (one example is shown in Figure~3),
but it is satisfying to analyze in detail
one particularly simple example based on the solution~\zs\ which itself
is already a special case of the most general dressed solution.

Therefore we look now at the
case $\lambda_1 = i$. Since the solution naively looks singular at
this value
we will carefully take the limit as $\lambda_1 \to i$ from inside
the unit circle.
To this end we consider
\eqn\limitone{
\lambda_1 = i a, \qquad
m = \sqrt{1 - a \over 1 + a}
}
in the limit $a \to 1$.
In this limit the equation for the boundary reduces to
\eqn\aaa{
\cosh U_1 = \sin U_2
}
whose solutions are just points in the $(U_1,U_2)$ plane.

In order to isolate what is going on near the point $(0,\pi/2)$
(for example) we should
rescale the worldsheet coordinates by defining
new coordinates $x, y$ according to
\eqn\aaa{
U_1 = 2 m x, \qquad U_2 = {\pi \over 2} + 2 m y
}
Then in the limit $a \to 1$ the equation becomes
\eqn\aaa{
0 = D = (1 - x^2 - y^2) (1 - a) + {\cal O}(1 - a)^2
}
So now the edge of the worldsheet is the circle
$x^2 + y^2 = 1$ in the $(x,y)$ plane.
Using \zzz\ and \yyy\ gives
\eqn\aaa{
u_1 = {1 \over 2} x (1-a), \qquad
u_2 = {\pi \over 4 a}\sqrt{1 - a^2} + {1 \over 2 a} (1 - a) y.
}
Plugging these values and~\limitone\ into the solution~\zs\ we can
then safely take $a \to 1$, obtaining
the surface
\eqn\aaa{
Z_1 = - i {1 + x^2 + y^2 \over 1 - x^2 - y^2}, \qquad
Z_2 = {2 i x - 2 y \over 1 - x^2 - y^2}.
}
Switching now to Poincar\'e coordinates
$(R,T,X)$ according to the usual embedding
\eqn\aaa{
Z_1 = {1 \over 2} \left({1 \over R} + {R^2 - T^2 + X^2 \over R}\right)
+ i {T \over R}, \qquad
Z_2 = {X \over R} + {i \over 2}\left(
{1 \over R} - {R^2 - T^2 + X^2 \over R} \right)
}
we find
\eqn\psol{
R =  {1 - x^2 - y^2 \over x}, \qquad
T = - {1 + x^2 + y^2 \over x}, \qquad
X = -{y \over x}.
}
Finally we note that this surface in $AdS_3$ satisfies
\eqn\aaa{
-T^2 + X^2 = -1 - {(1 - x^2 - y^2)^2 \over 4 x^2}.
}
At the edge of the worldsheet
the second term on the right-hand side is zero, so we conclude that
the solution~\psol\ intersects the boundary of $AdS_3$
along the curve described by
\eqn\aaa{
-T^2 + X^2 = - 1.
}
Interestingly this is a timelike curve whereas the giant gluon
solution we started with traces out a path of lightlike curves on the boundary.

More complicated cases must be studied numerically.
In Figure~3 we show the timelike curve on the boundary of $AdS_3$
that bounds the sample surface shown in Figure~2.

\fig{In this plot we consider, as an example, the solution~\zs\ for
the particular case $\lambda_1 = 1/2 + i/3$.
As explained in the text, the solution actually corresponds to
infinitely many Wilson loops on the boundary of $AdS_3$, one of which
is the curve shown here in the $(X,T)$
plane on the boundary of $AdS_3$ in Poincar\'e coordinates.
The light-cone to which these timelike curves
asymptote is also shown.  The Wilson loop is of course a closed curve;
the upper and lower branches shown here live on opposite sides of the
$AdS_3$ cylinder in global coordinates.
The minimal area surface spanning this curve is shown in Figure~2.
}{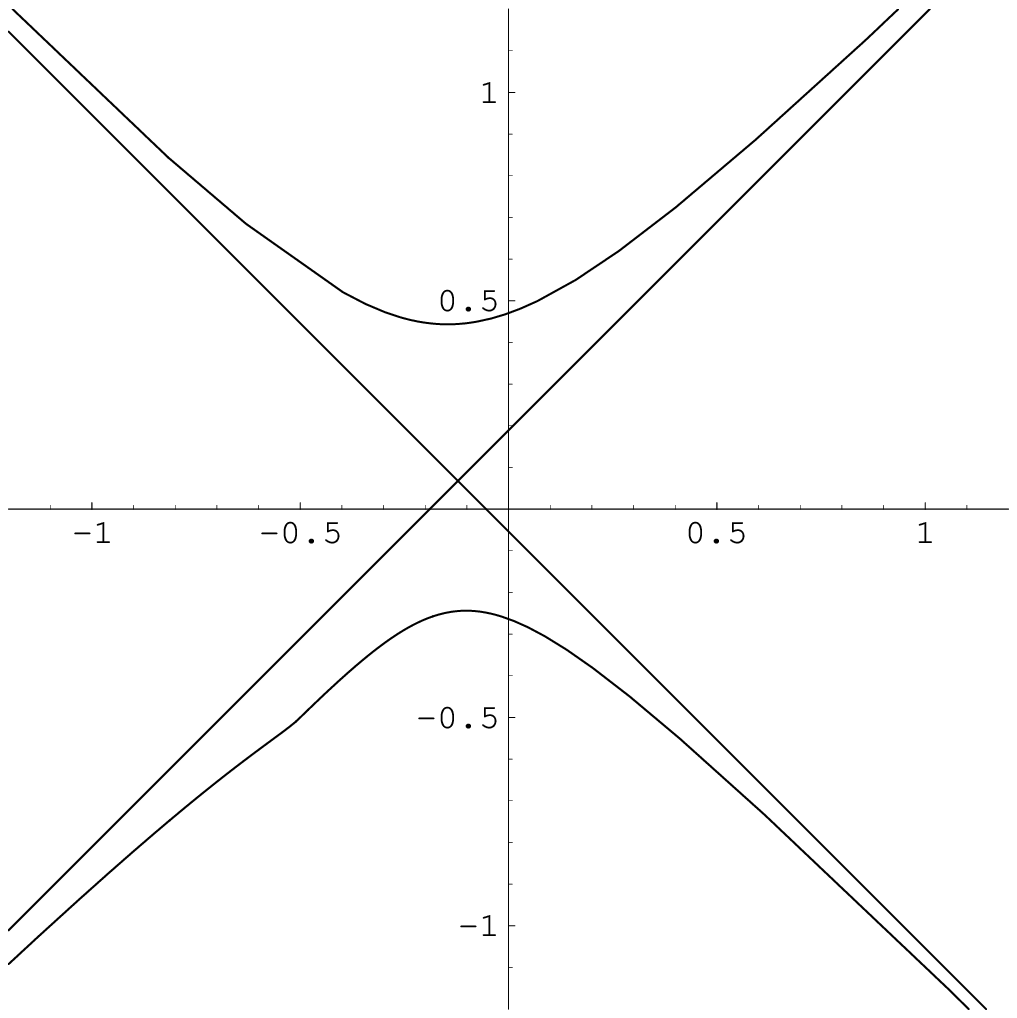}{2.0in}

\newsec{$AdS_5$ Solutions}

We now turn our attention to the dressing problem for worldsheets
in $AdS_5$.  This case is somewhat more complicated because it is
not realized as a principal chiral model.  Rather we use
the SU(2,2)/SO(4,1) coset model, parameterizing an element $g$ of the
coset 
in terms of the embedding coordinates $\vec{Y}$ according to~\ArutyunovYX
\eqn\fiveembed{
g = \pmatrix{
0 & +Z_1 & -Z_3 & +\bar{Z}_2 \cr
-Z_1 & 0 & +Z_2 & +\bar{Z}_3 \cr
+Z_3 & -Z_2 & 0 & -\bar{Z}_1 \cr
- \bar{Z}_2 & - \bar{Z}_3 & +\bar{Z}_1 & 0}
}
where
\eqn\zdefs{
Z_1 = Y_{-1} + i Y_0, \qquad
Z_2 = Y_1 + i Y_2,  \qquad
Z_3 = Y_3 + i Y_4.
}
This parameterization satisfies
\eqn\cosetcondition{
g^{\rm T} = - g, \qquad
g^\dagger M g = M,
}
where
\eqn\aaa{
M = \pmatrix{-1 & 0 & 0 & 0 \cr
0 & -1 & 0 & 0 \cr
0 & 0 & +1 & 0 \cr
0 & 0 & 0 & +1}
}
and has determinant
\eqn\aaa{
\det g = - \vec{Y} \cdot \vec{Y} = 1.
}
Taking again the giant gluon solution \solone\ (supplemented
with $Y_3 = Y_4 = 0$) as the `vacuum'
we now find that the solution to the linear system \linear\ is
\eqn\psifive{
\eqalign{
\Psi(\lambda) 
&=\left(\matrix{
0 & + \ch u_1 \ch U_2 + i m_- \sh u_1 \sh U_2 \cr
- \ch U_1 \ch u_2 - i m_+ \sh U_1 \sh u_2 & 0 \cr
0 & - \sh u_1 \ch U_2 - i m_- \ch u_1 \sh U_2 \cr
- m_+ \sh U_1 \ch u_2 + i \ch U_1 \sh u_2 & 0
}\right.
\cr
&\qquad\quad\quad
\left.\matrix{
0 & + m_- \sh u_1 \ch U_2 - i \ch u_1 \sh U_2 \cr
+ \sh U_1 \ch u_2 + i m_+ \ch U_1 \sh u_2 & 0 \cr
0 & - m_- \ch u_1 \ch U_2 + i \sh u_1 \sh U_2 \cr
+ m_+ \ch U_1 \ch u_2 - i \sh U_1 \sh u_2 & 0
}\right)
}}
in terms of
\eqn\intermsof{
U_1 = m_- z + m_+ \bar{z}, \qquad
U_2 = (m_- z - m_+ \bar{z})/i, \qquad
m_+ = 1/m_- = \left({1 + i \lambda \over 1 - i \lambda}\right)^{1/2}.
}
The solution \psifive\ has been chosen to satisfy the desired constraints
\eqn\aaa{
\Psi^\dagger(\bar{\lambda}) M \Psi(\lambda) = M, \qquad \det \Psi(\lambda) = 1
}
as well as the initial condition
\eqn\aaa{
\Psi(\lambda = 0) = g,
}
where $g$ is the giant gluon solution \solone\ 
written in the
embedding~\fiveembed.
Note that the symbols $U_1, U_2$ defined
in~\intermsof\ have been chosen because at $\lambda = 0$
they reduce to $u_1,u_2$.

\subsec{Construction of the dressing factor}

The dressing factor for this coset model takes the form
\eqn\chicoset{
\chi(\lambda) = 1 + {\lambda_1 - \bar{\lambda}_1 \over
\lambda - \lambda_1} P_1 + {1/\lambda_1 - 1/\bar{\lambda}_1
\over \lambda + 1/\bar{\lambda}_1} P_2.
}
In order to satisfy all the constraints on the dressed solution,
we choose $P_1$ and $P_2$ as follows.  First we choose $P_1$ to be
the hermitian (with respect to the metric $M$) projection operator
onto the vector $v_1 = \Psi(\bar{\lambda}_1) v$, where $v$ is an arbitrary
complex constant vector.
Specifically, $P_1$ is then given as in \projone\ by
\eqn\pone{
P_1 = { v_1 v_1^\dagger M \over v_1^\dagger M v_1},
}
which satisfies
\eqn\usdfh{
P_1^2 = P_1, \qquad P_1^\dagger = M P_1 M
}
as desired.
Next we choose
\eqn\choicetwo{
P_2 = \Psi(0) P_1^{\rm T} \Psi(0)^{-1}.
}
Because of \usdfh\ it is easy to check that $P_2$ also satisfies
\eqn\aaa{
P_2^2 = P_2, \qquad P_2^\dagger = M P_2 M,
}
so $P_2$ is also a hermitian projection operator; in fact it is easy
to check that $P_2$ projects onto the vector
\eqn\aaa{
v_2 = \Psi(0) M \overline{v_1}
}
and hence can be written as
\eqn\aaa{
P_2 = {v_2 v_2^\dagger M \over v_2^\dagger M v_2}.
}
Now let us explain the choice \choicetwo.  Notice that
\eqn\aaa{
v_2^\dagger M v_1 = v_1^{\rm T} M \Psi(0)^\dagger M v_1
= v_1^{\rm T} \Psi(0)^{-1} v_1
}
where we used $\Psi(0)^\dagger M \Psi(0) = M$.  But since
$\Psi(0)$ is antisymmetric, this is zero!  So $v_2$ and $v_1$
are orthogonal, and hence
\eqn\aaa{
P_1 P_2 = P_2 P_1 = 0.
}

Using all of the above relations one can check that \chicoset\ satisfies
the conditions
\eqn\aaa{
[\chi(\bar{\lambda})]^\dagger M \chi(\lambda) = M,
\qquad
\Psi^{\rm T}(0) \chi^{\rm T}(0) = - \chi(0) \Psi(0),
}
which guarantee that the dressed solution $\Psi'(\lambda) = \chi(\lambda)
\Psi(\lambda)$ continues to satisfy \cosetcondition.
As in the $AdS_3$ case we find that $\chi$ does not have unit determinant
but rather
\eqn\aaa{
\det \chi(\lambda) =  {\lambda - \bar{\lambda}_1 \over \lambda - \lambda_1}
{ \lambda - 1/\lambda_1 \over
\lambda - 1/\bar{\lambda}_1}.
}
We must therefore rescale the dressed solution
$\Psi'(0) = \chi(0) \Psi(0)$
by a factor of $\sqrt{\lambda_1/\bar{\lambda}_1}$.

To summarize, the dressed solution $g'$ is given by
\eqn\aaa{
g' = \sqrt{\lambda_1 \over \bar{\lambda}_1} \left[
1 + { \lambda_1 - \bar{\lambda}_1 \over - \lambda_1} P_1
+ { 1/\lambda_1 - 1/\bar{\lambda}_1 \over 1/\bar{\lambda_1}} P_2
\right] \Psi(0)
}
in terms of~\psifive\ and the projection operators~\pone, \choicetwo.
The solution is characterized by an arbitrary complex parameter
$\lambda_1$ and the choice of a complex four-component vector~$v$.

\subsec{A special case}

Since the general solution is again rather complicated we display
only a special case, choosing the vector $v = \pmatrix{1 & i & 0 & 0}$.
We then find
that the dressed solution $g'$ has the form~\fiveembed\ with
\eqn\aaa{
Z'_1 = {1 \over |\lambda_1|}{\vec{Y} \cdot \vec{N}_1 \over D},
\qquad Z'_2 ={1 \over |\lambda_1|}
 {\vec{Y} \cdot \vec{N}_2 \over D}, \qquad
Z'_3={1 \over |\lambda_1|}{N_3 \over D}
}
in terms of the numerator factors
\eqn\aaa{\eqalign{
\vec{N}_1 &=
 \pmatrix{-|m|^2 \bar{\lambda}_1 (\ch U_1 \ch \bar{U}_1+\ch U_2 \ch \bar{U}_2)+\lambda_1 \sh U_1 \sh \bar{U}_1 +|m|^4 \lambda_1 \sh U_2 \sh \bar{U}_2  \cr
 -i |m|^2 \lambda_1 (\ch U_1 \ch \bar{U}_1+\ch U_2 \ch \bar{U}_2)+i \bar{\lambda}_1\sh U_1 \sh \bar{U}_1+i|m|^4 \bar{\lambda}_1\sh U_2 \sh \bar{U}_2 \cr
 -(\lambda_1-\bar{\lambda}_1) \bar{m} \sh U_1 \ch \bar{U}_1+i(\lambda_1-\bar{\lambda}_1)\bar{m} |m|^2  \ch U_2 \sh \bar{U}_2 \cr
 +i (\lambda_1-\bar{\lambda}_1) m \ch U_1 \sh \bar{U}_1-(\lambda_1-\bar{\lambda}_1) m |m|^2 \sh U_2 \ch \bar{U}_2},\cr
 \vec{N}_2 &=  \pmatrix{-(\lambda_1-\bar{\lambda}_1) \bar{m}\sh U_1 \ch \bar{U}_1+i(\lambda_1-\bar{\lambda}_1) \bar{m} |m|^2  \ch U_2 \sh \bar{U}_2 \cr
 +i (\lambda_1-\bar{\lambda}_1) m \ch U_1 \sh \bar{U}_1-(\lambda_1-\bar{\lambda}_1) m |m|^2 \sh U_2 \ch \bar{U}_2 \cr
 -|m|^2 \bar{\lambda}_1 (\ch U_1 \ch \bar{U}_1+\ch U_2 \ch \bar{U}_2)+\lambda_1 \sh U_1 \sh \bar{U}_1 +|m|^4 \lambda_1 \sh U_2 \sh \bar{U}_2\cr
 -i |m|^2 \lambda_1 (\ch U_1 \ch \bar{U}_1+\ch U_2 \ch \bar{U}_2)+i \bar{\lambda}_1 \sh U_1 \sh \bar{U}_1+i |m|^4 \bar{\lambda}_1\sh U_2 \sh \bar{U}_2},\cr
  N_3 &=\bar{m} (\lambda_1-\bar{\lambda}_1) (-i \sh U_1 \ch \bar{U}_2+|m|^2 \ch U_1 \sh \bar{U}_2),
}}
$\vec{Y}$ again given in~\solone,
and
the denominator
\eqn\aaa{
D = -|m|^2 (\ch U_1 \ch \bar{U}_1 +\ch U_2 \ch \bar{U}_2)+\sh U_1 \sh \bar{U}_1+|m|^4 \sh U_2 \sh \bar{U}_2.
}
In these expressions $m$ and $\bar{m}$ are as in~\mmbardefs, with
\eqn\aaa{
U_1 = z/m + m \bar{z}, \qquad U_2 = (z/m - m \bar{z})/i.
}
The real embedding coordinates $\vec{Y}'$ of the dressed
solution may then be extracted 
from~\zdefs.
It is straightforward, though somewhat tedious, to directly verify
that the resulting $\vec{Y}'$
satisfies the equations of motion~\eom\ and the Virasoro
constraints~\virasoro, providing a check on our application of
the dressing method.

\bigskip

\centerline {\bf Acknowledgments}

\smallskip

We are grateful to J.~Maldacena for comments.
The work of AJ is supported in part by the Department of Energy
under contract DE-FG02-91ER40688.
The work of MS is supported by NSF grant PHY-0610259 and by an OJI award
under DOE contract DE-FG02-91ER40688.
The work of AV is supported by NSF CAREER Award PHY-0643150.

\appendix{A}{Conventions}

Here we summarize the standard conventions
for global $AdS_3$ that we have used in preparing Figures~1 and~2.
We parametrize the SU(1,1) group element~\parametrization\ as
\eqn\aaa{
g = \pmatrix{
e^{+i \tau} \sec \theta & e^{+i \phi} \tan \theta \cr
e^{-i \phi} \tan \theta & e^{-i \tau} \sec \theta},
}
where $\tau$ is global time, $\phi$ is the azimuthal angle, 
and $\theta$ runs from $0$ in the interior of the $AdS_3$ cylinder
to $\pi/2$ at the boundary of $AdS_3$.
In terms of these quantities the parametric plots in Figures~1 and~2
have Cartesian coordinates
\eqn\aaa{
(x,y,z) = ( \theta \cos \phi, \theta \sin \phi, \tau)
}
and the boundary of $AdS_3$ is the cylinder $x^2 + y^2 = (\pi/2)^2$.

\listrefs

\end